\documentclass[preprint,
superscriptaddress,
preprint,
 amsmath,amssymb,
aps,
pre,
]{revtex4-1}

\usepackage{amssymb}
\usepackage{amsthm}
\usepackage{graphicx}
\usepackage{dcolumn}
\usepackage{bm}
\usepackage{amsfonts}
\usepackage{amsmath}
\usepackage{amssymb}
\usepackage{color}
\usepackage{natbib}
\usepackage{epstopdf}
\usepackage{soul}

\begin{document}

\title{Oscillations of small bubbles and medium yielding in elastoviscoplastic fluids\\} 

\author{Marco De Corato} 
\address{Department of Chemical Engineering, Imperial College London, London, SW7 2AZ, United Kingdom}
\address{Current address: Institue for Bioengineering of Catalonia (IBEC), Baldiri i Reixac 10-12, 08028 Barcelona, Spain}
\author{Brice Saint-Michel }%
\address{Department of Chemical Engineering, Imperial College London, London, SW7 2AZ, United Kingdom}
\author{George Makrigiorgios}
\address{Fluid Mechanics and Rheology Laboratory, Department of Chemical Engineering, University of Patras, Patras 26500, Greece}
\address{Current address: Department of Chemical and Biomolecular Engineering, University of California, Berkeley, CA 94720}
\author{Yannis Dimakopoulos }
\address{Fluid Mechanics and Rheology Laboratory, Department of Chemical Engineering, University of Patras, Patras 26500, Greece}
\author{John Tsamopoulos }
\address{Fluid Mechanics and Rheology Laboratory, Department of Chemical Engineering, University of Patras, Patras 26500, Greece}
\email{tsamo@chemeng.upatras.gr}%
\author{Valeria Garbin } 
\address{Department of Chemical Engineering, Imperial College London, London, SW7 2AZ, United Kingdom}
\email{v.garbin@imperial.ac.uk}


\date{\today}

\begin{abstract}
We investigate the radial oscillations of small gas bubbles trapped in yield-stress fluids and driven by an acoustic pressure field. We model the  rheological behavior of the yield-stress fluid using the recently developed elasto-visco-plastic (EVP) constitutive equation that takes into account the elastic and visco-plastic deformations of the material [P. Saramito, J. NonNewton. Fluid Mech. 158 (1-3) (2009) pp. 154–161]. Assuming that the bubble remains spherical during the pressure driving, we reduce the problem to a set of ODEs and an integrodifferential equation, which we solve numerically for the case of two yield stress fluids, a soft Carbopol gel and a stiffer Kaolin suspension. We find that, depending on the amplitude and frequency of the pressure field, the radial oscillations of the bubble produce elastic stresses that may or may not suffice to yield the surrounding material. We evaluate the critical amplitude of the acoustic pressure required to achieve yielding and we find a good agreement between numerical simulations and an analytical formula derived under the assumption of linear deformations.
Finally, we examine the bubble oscillation amplitude for a very wide range of applied pressures both below and above the critical value to assess the impact of yielding on the bubble dynamics.
This analysis could be used to identify a signature of yielding in experiments where the radial dynamics of a bubble is measured. More generally, these results can be used to rationalize the optimal conditions for pressure-induced bubble release from yield-stress fluids, which is relevant to various biomedical and industrial applications, including oil industry and food processing.
\end{abstract}

\maketitle






\section{introduction}
\label{sec:introduction}
Yield-stress fluids encompass a large variety of soft materials, e.g. pastes, slurries, emulsions and microgels which possess a characteristic stress $\tau_{y}$ below which they stop flowing and behave as solids~\cite{coussot2014yield,bonn2017yield}. They may trap bubbles when buoyancy-induced stresses do not suffice to yield the material~\cite{beris1985creeping,dubash2004conditions,dubash2007propagation,singh2008interacting,tsamopoulos2008steady,sikorski2009motion,dimakopoulos2013steady,tripathi2015bubble,lopez2018rising}. 

Significant bubble entrapment is desirable in food engineering to improve texture and slow down melting of ice cream~\cite{sofjan2004effects} and improve the perceived crunchiness of snacks~\cite{luyten2004crispy}. A small, controlled amount of bubbles is also allowed in concrete to improve workability and freeze-thaw resistance, at the expense of reduced compressive strength and concrete blisters~\cite{kosmatka2011design}.
Bubble entrapment in yield-stress fluids can be an undesirable byproduct of fluid processing: the presence of air pockets in polydimethylsiloxane (PDMS) used for microfluidics applications can severely reduce its quality and transparency~\cite{mazzeo2009centrifugal}; bubbles trapped in fluids used in the oil industry can result in undesired permeable slurries~\cite{bonett1996getting} or lead to explosions in drilling mud, which may delay production and potentially inflict huge burden on the ecosystem \cite{Deep_water,johnson1991gas}. Air bubbles induce bacterial contamination and poor final appearance in personal care products~\cite{lin1970mechanisms}. 

Complete bubble removal can be achieved, by centrifuging the sample or by using a vacuum pump to inflate bubbles. Recent experiments have confirmed that mechanical agitation is an effective method of removing gas voids from granular materials for which the yield stress is a consequence of the frictional contact network formed by the microscopic phase \cite{koch2019vibration}. 
Stein and Buggisch~\cite{stein2000rise} and Karapetsas et al. \cite{Karapetsas2018} showed that driving a bubble into volumetric oscillations using a dynamic external pressure can generate sufficiently large deformation and mechanical stresses to locally yield the material, thus promoting bubble release from yield-stress fluids. The same mechanism has been exploited by Iwata et al.~\cite{iwata2008pressure,IWATA201924} to enhance bubble removal from highly viscous shear-thinning and viscoelastic fluids.

Understanding bubble dynamics in complex fluids is then paramount to achieve controlled bubble release. In viscoelastic media, the interplay between the bubble shape and the rheological response of the fluid in bubble rise experiments are now well understood~\cite{astarita1965motion,hassager1979negative,pilz2007critical,papaioannou2014bubble,fraggedakis2016velocity} whereas the abundant literature on acoustically-driven bubble oscillation revealed delayed collapse~\cite{ellis1970cavitation,fogler1970collapse,papanastasiou1984bubble,kim1994collapse} and chaotic bubble oscillations~\cite{jimenez2005bubble,naude2008periodic,kafiabad2010chaotic,cunha2013oscillatory}. Focus has progressively shifted from the initial context of damage to military ships  towards damage in soft tissues for biomedical applications~\cite{allen2000dynamics,allen2000dynamics1,holt2001measurements,yang2005model,hua2013nonlinear,warnez2015numerical,gaudron2015bubble,movahed2016cavitation,zilonova2018bubble} and high-frequency rheology of soft materials ~\cite{hamaguchi2015linear,jamburidze2017high,estrada2018high}, as recently reviewed by Dollet et al.~\cite{garbin2019}.

In contrast, the literature on bubble dynamics in yield-stress fluids is very limited and focuses almost exclusively on the problem of bubble rise due to buoyancy. In such fluids, rising motion results from the interplay between the gravity, bubble shape, the position of the yield surface and the rheology of the material both below and above the yield stress. Experiments require great care to suppress internal stresses and achieve repeatability~\cite{mougin2012significant, lopez2018rising}. Numerical investigation of the problem proves to be equally challenging. The existing works model the material rheology using either Bingham or Herschel-Bulkley constitutive equations~\cite{yang1966theoretical,stein2000rise,singh2008interacting,tsamopoulos2008steady,dimakopoulos2013steady,tripathi2015bubble}. 
Both these models predict a discontinuity of the viscosity at the yield surface, the location of which is unknown in flows that are 2D, 3D or time-dependent. To avoid this problem and make numerical solutions feasible these equations are either regularized (e.g. \cite{papanastasiou1987flows}), which unfortunately reduces the solid region of the material to a liquid with very large viscosity, or solved via the Augmented Lagrangian Method~\cite{dimakopoulos2013steady}, which converges extremely slowly, but recently has been improved \cite{dimakopoulos2018pal}.
More importantly, the Bingham and Herschel-Bulkley models implicitly assume that the unyielded material cannot deform, even if the yield surface may adjust to flow, especially in time-dependent problems, and leave undetermined the stress field there. While this problem is not critical for the case of bubble rise, bubble oscillation prescribes a non-zero strain field everywhere in the fluid. This leads to two unphysical results: first, it implies that any finite oscillation amplitude causes yielding of all the material; second, it means that the stress applied by the bubble is everywhere above the yield stress, and is therefore infinite. The former issue contrasts with the experimental findings of Stein and Buggisch~\cite{stein2000rise} who report that a finite oscillation amplitude is required to achieve yielding. This issue is not discussed in \cite{yang1966theoretical} and is circumvented in \cite{stein2000rise} by prescribing the deformation field and thus the dynamics of the yield-surface. Recently, Karapetsas et al. \cite{Karapetsas2018} performed a detailed theoretical and numerical analysis of a bubble rising in a Bingham fluid and driven by an acoustic field into volumetric oscillations. The authors developed a simplified 1D spherosymmetric model and also performed detailed numerical simulations that take into account axisymmetirc deformations of the gas-liquid interface. Their results confirm that a Bingham material is yielded everywhere during the oscillations of the bubble, that a yield-surface cannot be defined for this type of flow and that an acoustic field promotes the release of bubbles that would be trapped by the yield stress otherwise.

In this paper, we investigate the radial oscillations of a microbubble trapped in an elastic yield-stress fluid and driven by an external pressure field. We focus on the case of ultrasonic fields in the frequency range $\sim 10-100 \, \rm{kHz}$, which is relevant to industrial equipment. Small bubbles with radii $\sim 30-300 \, \rm{\mu m}$ are resonant in this frequency range, that is, they are efficiently excited into volumetric oscillations by the ultrasound field. 
We perform numerical simulations employing a generalized Rayleigh-Plesset equation and a recently developed elasto-visco-plastic (EVP) constitutive equation that takes into account elastic and visco-plastic deformations of the material \cite{saramito2009new}. Employing this model we resolve the conceptual difficulties introduced in~\cite{yang1966theoretical,stein2000rise} by the choice of the Bingham model. Using numerical simulations and an approximate linear theory, we evaluate the critical acoustic pressure required to yield the material and compute the dynamics of the yield surface. Finally, we explore the impact of yielding on the radial oscillations of the bubble. Our results represent a first step towards the investigation of pressure-induced bubble release from yield-stress fluids and could be potentially used to identify the signature of yielding in experiments. The theoretical and numerical framework developed in this paper can be applied to validate constitutive equations for yield-stress fluids by comparing with experiments performed under controlled extensional deformation imparted via bubble oscillation.

\section{Equations governing the bubble dynamics}
\label{sec:governing_equations}
We consider a bubble of equilibrium radius $R_0$ suspended in an incompressible yield-stress fluid. The bubble is driven by a time-dependent pressure $p_\infty (t)$ imposed far from the bubble.
We assume that the Cauchy stress tensor, $\boldsymbol{T}$, is given by:
\begin{equation} \label{cauchy}
\boldsymbol{T} = -p \boldsymbol{I} + \eta_s \left(\nabla \boldsymbol{v}+\nabla \boldsymbol{v}^T\right)+\boldsymbol{\tau} \, \, \, ,
\end{equation}
where $p$ denotes the pressure, $\boldsymbol{v}$ is the velocity field, $\eta_s$ is the shear viscosity of the solvent and $\boldsymbol{\tau}$ is a non-traceless and non-Newtonian contribution to the total stress.
In the present work we use the elasto-viscoplastic model developed by Saramito \cite{saramito2009new} that gives excellent predictions when compared to experimental results in shear \cite{fraggedakis2016yielding} and for the case of a sedimenting sphere \cite{fraggedakis2016yielding1}. This model predicts a Neo-Hookean elastic behavior before yielding and a viscoelastic behavior afterwards, with an instantaneous transition from solid- to liquid-like behavior when the second invariant of the deviatoric part of $\boldsymbol{\tau}$ is larger than the yield stress.  
The evolution of $\boldsymbol{\tau}$ is governed by:
\begin{equation}\label{model1}
\frac{1}{G}\overset{\triangledown}{\boldsymbol{\tau}} = \nabla \boldsymbol{v}+\nabla \boldsymbol{v}^T -\text{max} \left[0, \frac{\left\vert \boldsymbol{\tau}_d \right\vert-\tau_y}{K}\right]^{1/n} \, \frac{\boldsymbol{\tau}}{\left\vert \boldsymbol{\tau}_d \right\vert} \, \, \, ,
\end{equation}
where $G$ is the elastic modulus of the material and the symbol $\triangledown$ above a tensor $\boldsymbol{X}$ denotes its upper convected derivative, defined as: 
\begin{equation}
 \overset{\triangledown}{\boldsymbol{X}}= \frac{D \boldsymbol{X}}{D t}  -\nabla \boldsymbol{v}^T \cdot \boldsymbol{X}-\boldsymbol{X} \cdot \nabla \boldsymbol{v} \, \, \, ,
\end{equation}
with $D/D t$ the material derivative. In Eq. \eqref{model1}, $\left\vert \boldsymbol{\tau}_d \right\vert$ denotes the square root of the second invariant of the deviatoric part of the stress tensor $\boldsymbol{\tau}$, and $K$ is the consistency parameter of the yielded phase with $n$ its power-law index. For stresses  $\left\vert \boldsymbol{\tau}_d \right\vert < \tau_y$ the material is unyielded and experiences no viscoplastic deformation, for $\left\vert \boldsymbol{\tau}_d \right\vert > \tau_y$ the material behaves as a viscoelastic fluid thus undergoing both elastic and viscoplastic deformation. In the limit $G \rightarrow \infty$, the constitutive model given by Eq. \eqref{model1} reduces to the well-known Herschel-Bulkley model. In the linear regime, Eqs. \eqref{cauchy}-\eqref{model1} reduce to the Kelvin-Voigt solid model. Important advantages of this model (especially in comparison to the Bingham and Herschel-Bulkley models) are that  it determines the flow and stress fields in both material regions, does not require regularization and converges quite fast to the solution.
Finally, the Von Mises criterion describes the critical stress state above which the material starts to experience viscoplastic flow in Eq. \eqref{model1}, in agreement with experimental observation in the case of multiaxial deformations \cite{ovarlez2010three, martinie2013apparent}. However, recent experiments considering the extensional deformation of yield-stress fluids \cite{martinie2013apparent,zhang2018yielding} suggest that the third invariant of $\boldsymbol{\tau}_d$ may be also required in the yielding criterion.

Given the small size of the bubbles examined ($\sim 100 \, \rm{\mu m}$), the relevant Bond number is less than $10^{-3}$, hence we assume that the center of volume of the bubble is stationary. We define a spherical coordinate system with its origin at the center of the bubble, with $r$, $\theta$ and $\phi$ the radial, azimuthal and polar coordinates, respectively. Under the assumption that the bubble undergoes spherical oscillations only, the problem is spherically symmetric, implying: 
\begin{subequations}\label{spherical_symm}
\begin{alignat}{1}
\! & \boldsymbol{v}=v_r(r) \, \boldsymbol{\hat{r}} \, \, \, , \\
\! & \tau_{\theta \theta}=\tau_{\phi \phi} \, \, \, , \\
\! & \boldsymbol{\tau} = \tau_{rr}(r) \, \boldsymbol{\hat{r}}\boldsymbol{\hat{r}} +\tau_{\theta \theta}(r) \, \boldsymbol{\hat{\theta}}\boldsymbol{\hat{\theta}}+\tau_{\phi \phi}(r) \, \boldsymbol{\hat{\phi}}\boldsymbol{\hat{\phi}} \, \, \, , \\
\! & p = p(r) \, \, \, .
\end{alignat}
\end{subequations}
The incompressibility of the material implies that the radial velocity, $v_r$, is related to the time-dependent radius of the bubble, $R$, through:
\begin{equation}\label{rad_vel}
v_r = \frac{R^2 \dot{R}}{r^2} \, \, \, .
\end{equation}
In the case of $\dot{R}>0$ the bubble expands and the material undergoes a non-uniform spherosymmetric compression. A nonuniform spherosymmetric extension is applied to the medium during bubble compression $\dot{R}<0$.
We assume that the pressure at infinity changes due to the acoustic driving as $p_\infty(t)$. The time evolution of the bubble radius is governed by the generalized Rayleigh-Plesset equation that is obtained by integrating the radial component of the momentum balance from $R$ to infinity \cite{prosperetti1982generalization}:
\begin{equation}\label{genRP1}
\rho \left(\ddot{R}R +\frac{3}{2} \dot{R}^2 \right) =  p(R) -p_\infty(t)-\tau_{rr}(R) + 2 \int_R^\infty  \frac{\tau_{rr}-\tau_{\theta \theta}}{r} dr \, \, \, ,
\end{equation}
where $\rho$ is the density of the medium and we have assumed that the stress tensor $\boldsymbol{\tau}$ vanishes at infinity because the rate of strain goes to zero far from the bubble. In Eq. \eqref{genRP1},  $ p(R)$ and $\tau_{rr}(R)$ are the pressure and the radial component of $\boldsymbol{\tau}$ evaluated at the surface of the bubble, respectively. These quantities are related through the normal stress balance:
\begin{equation}
p(R) = p_{\text{gas}} +\tau_{rr}(R) - \eta_s \frac{4 \dot{R}}{R} - \frac{2 \gamma}{R} \, \, \, ,
\end{equation}
where $\gamma$ is the surface tension of the interface between the gas and the yield-stress fluid. In principle the surface tension between the gas and the yield-stress fluid could depend on the rheological state of the material. In this paper we assume that the surface tension is the same, regardless of the stresses inside the yield-stress fluid. We denote the pressure inside the bubble with $p_{\text{gas}}$ and we neglect the viscous stresses in the gas phase. We assume that the bubble undergoes isothermal compression $p_{\text{gas}} = \left( p_0 + 2\gamma/R_0 \right) (R_0/R)^3$ and that the driving pressure is periodic $p_\infty(t) = p_0 + \Delta p \, \sin{\left(\omega \, t\right)}$, with the equilibrium pressure $p_0$, the driving pressure amplitude $\Delta p$ and the angular frequency $\omega$.

The dimensionless form of Eqs. \eqref{genRP1} and \eqref{model1} are obtained introducing the following dimensionless quantities, noted with stars $*$:
\begin{equation}
t^* = \omega t , \;\;\;  R^* = \frac{R}{R_0}, \;\;\;  \boldsymbol{v}^* = \frac{\boldsymbol{v}}{R_0 \omega} , \;\;\; \boldsymbol{\tau}^* = \frac{\boldsymbol{\tau}}{K \omega^n} , \;\;\;  p^* = \frac{p}{K \omega^n} \; .
\end{equation}
Substituting the pressure $p(R)$ in Eq. \eqref{genRP1} and dropping the stars we obtain:
\begin{equation}\label{genRP_nd}
Re \left(\ddot{R}R +\frac{3}{2} \dot{R}^2 \right) = P_0\left[ \frac{\left( 1 + P_\text{stat}^{-1}\right)}{R^3} - \frac{P_\text{stat}^{-1}}{R} -1 \right]-\alpha \frac{ \dot{R}}{R} - P \, \sin{\left(t\right)}+ 2 \int_R^\infty  \frac{\tau_{rr}-\tau_{\theta \theta}}{r} dr \, \, \, ,
\end{equation}
\begin{subequations}\label{nondimens_eqs}
\begin{alignat}{1}
\! & De \left( \frac{ \partial \tau_{rr}}{\partial t} + \frac{R^2 \dot{R}}{r^2} \frac{\partial \tau_{rr}}{\partial r} + \frac{4 R^2 \dot{R}}{r^3} \tau_{rr} \right) = - \frac{4 R^2 \dot{R}}{r^3}-\text{max} \left[0, \frac{\left\vert \tau_{rr} -\tau_{\theta \theta} \right\vert-\sqrt{3} Bn}{\sqrt{3}}\right]^{1/n} \, \frac{ \sqrt{3}\tau_{rr}}{\left\vert \tau_{rr} -\tau_{\theta \theta} \right\vert} \, \, , \\
\! & De \left( \frac{\partial \tau_{\theta \theta}}{\partial t} + \frac{R^2 \dot{R}}{r^2} \frac{\partial \tau_{\theta \theta}}{\partial r} - \frac{2 R^2 \dot{R}}{r^3} \tau_{\theta \theta} \right) = \frac{2 R^2 \dot{R}}{r^3}-\text{max} \left[0, \frac{\left\vert \tau_{rr} -\tau_{\theta \theta} \right\vert-\sqrt{3} Bn}{\sqrt{3}}\right]^{1/n} \, \frac{ \sqrt{3}\tau_{\theta \theta}}{\left\vert \tau_{rr} -\tau_{\theta \theta} \right\vert} \, \, .
\end{alignat}
\end{subequations}
In Eqs. \eqref{genRP_nd}-\eqref{nondimens_eqs} we have introduced the relevant dimensionless numbers. The ratio of the solvent to the  viscosity of the yield stress material is given by $\alpha = \eta_s \omega^{1-n}/K$. The Bingham number, $Bn=  \tau_y/ K \omega^n$, denotes the ratio of the yield stress to the material viscous stress. The Reynolds number, $Re= \rho R_0^2 \omega^{2-n}/K$, expresses the relative importance of inertial to viscous stresses. The dimensionless number $P_\text{stat} = R_0 p_0/ 2 \gamma $ gives the ratio of the static pressure to surface tension. The Deborah number, $De= K \omega^n/G$ expresses the ratio of the viscoelastic relaxation time to the characteristic flow timescale. The dimensionless ambient pressure is given by $P_0 = p_0/ K \omega^n$ and the dimensionless pressure amplitude, $\Delta P = \Delta p/K \omega^n$, compares the amplitude of acoustic pressure to the viscous stresses. 

\begin{table}
\begin{tabular}{ l c c c c c c r }
  \hline			
  $\text{Fluid}$ & $\tau_y \, [\rm{Pa}] $& $K \, [\rm{Pa} \, \rm{s}^\textit{n} ]$ & $n$ & $G \, [\rm{Pa}]$ & $\rho \, [\rm{kg \, m^{-3}}]$ & $\gamma \, [\rm{Pa \, m} ]$ & $\text{ref}$ \\ \hline
  \text{Carbopol} & 8.6 & 3.5 & 0.43 & 81.5 & 1000 & 0.07 & \text{\cite{lacaze2015steady}} \\
  \text{Kaolin} & 91 & 68 & 0.39 & 200000 & 1630 & 0.07 & \text{\cite{luu2009drop}} \\
  \hline  
\end{tabular}
\caption[Table 1:]{Properties of the yield-stress fluids.}\label{table:table1}
\end{table}

The values of the constitutive parameters for yield-stress fluids can vary over a wide range. In this work, we consider two yield-stress fluids: a soft Carbopol gel and a stiff Kaolin suspension. The physical properties of the two materials are summarized in Table \ref{table:table1} and are taken from the measurements of Lacaze et al. \cite{lacaze2015steady} and Luu and Forterre \cite{luu2009drop}. Lacaze et al. used a Carbopol 940 gel at a concentration of $0.1 \%$ in weight. The Kaolin used by Luu and Forterre is a colloidal suspension of clay in water at $55 \%$ in weight, which was supplied by Imerys Ceramics France. Both yield-stress fluids display a pronounced shear-thinning response. The large difference in the elastic moduli between the two fluids allows us to explore different regimes of the bubble dynamics. Since the solvent used in the Carbopol gel and in the Kaolin suspension is water, we assume that the Newtonian viscosity in Eq. \eqref{cauchy} is given by $\eta_s=0.001 \, \rm{Pa \, s}$. This makes $\alpha$ small, typically less than $0.01$. In appendix A, we report the rheological response predicted by the EVP constitutive equation for the two yield-stress fluids.
Finally, throughout this paper we fix the ambient pressure $p_0 = 1.13 \, \times 10^5 \, \rm{Pa}$. We neglect residual stresses that are potentially present in the material after its preparation and that decay over long timescales \cite{mougin2012significant, dinkgreve2018carbopol}. Thus, we assume equilibrium initial conditions: $R(0)=R_0$, $\dot{R}(0)=0$, and $\tau_{rr}(r,0) = \tau_{\theta \theta}(r,0) = 0$.

\section{Numerical approach}
The solution of the partial differential equations \eqref{genRP_nd}-\eqref{nondimens_eqs} in the present form is complicated by the motion of the bubble surface that makes the domain time-dependent. To avoid this problem and immobilize the boundary we follow previous works \cite{zana1975dissolution, allen2000dynamics} and transform the radial coordinate into the Lagrangian coordinate:
\begin{equation}
y = r^3-R^3 \, \, \, .
\end{equation}
This coordinate transformation reduces Eqs. \eqref{genRP_nd}-\eqref{nondimens_eqs} to a system of first-order ordinary differential and integro-differential equations defined on $y$, where the surface of the bubble is given by the point $y=0$:
\begin{equation}\label{genRP_nd_Lag}
\frac{d R}{d t} = U
\end{equation}
\begin{equation}\label{genRP_nd_Lag1}
Re \left(\frac{dU}{dt} R +\frac{3}{2} U^2 \right) = P_0 \left[ \frac{\left( 1 + P_\text{stat}^{-1} \right)}{R^3} - \frac{P_\text{stat}^{-1}}{R} -1 \right]-\alpha \frac{U}{R} - P \, \sin{\left(t\right)}+ \frac{2}{3} \int_0^\infty  \frac{\tau_{rr}-\tau_{\theta \theta}}{y+R^3} dy \, \, \, ,
\end{equation}
\begin{subequations}\label{nondimens_eqs_Lag}
\begin{alignat}{1}
\! & De \left( \frac{d\tau_{rr}}{dt} +  \frac{4 R^2 U}{y+R^3} \tau_{rr} \right) = - \frac{4 R^2 U}{y+R^3}-\text{max} \left[0, \frac{\left\vert \tau_{rr} -\tau_{\theta \theta} \right\vert-\sqrt{3} Bn}{\sqrt{3}}\right]^{1/n} \, \frac{ \sqrt{3}\tau_{rr}}{\left\vert \tau_{rr} -\tau_{\theta \theta} \right\vert} \, \, , \\
\! & De \left( \frac{d\tau_{\theta \theta}}{dt} - \frac{2 R^2 U}{y+R^3} \tau_{\theta \theta} \right) =  \frac{2 R^2 U}{y+R^3}-\text{max} \left[0, \frac{\left\vert \tau_{rr} -\tau_{\theta \theta} \right\vert-\sqrt{3} Bn}{\sqrt{3}}\right]^{1/n} \, \frac{ \sqrt{3}\tau_{\theta \theta}}{\left\vert \tau_{rr} -\tau_{\theta \theta} \right\vert} \, \, .
\end{alignat}
\end{subequations}
We discretize the coordinate $y$ into a set of $N$ points $y_i = y_1,...,y_N$, with the last point describing the conditions far from the bubble. Following Kafiabad and Sadeghy \cite{kafiabad2010chaotic}, we discretize the spatial integral in Eq. \eqref{genRP_nd_Lag1} using a Gauss-Laguerre quadrature method, which is suitable for integrals of the type $\int_0^\infty e^{-x}f(x)dx$. To apply the Gauss-Laguerre method we first rewrite the spatial integral in Eq. \ref{genRP_nd_Lag1} as $I = \int_0^\infty e^{-y} \left[ e^y f(y) \right] dy$, with $f(y)=\left(\tau_{rr}-\tau_{\theta \theta}\right)/\left(y+R^3\right)$ and then we approximate $I$ with the sum 
\begin{equation}
I \approx \sum_{i=1}^N w_i e^{y_i}f(y_i) \, \, \, ,
\end{equation}
where $w_i$ are the weight factors defined as $w_i = y_i/\left[\left(L_{N+1}\left(y_i\right)\right)^2 \left(N+1\right)^2\right]$ and $L_{N+1}(y_i)$ is the Laguerre polynomial of order $N+1$ evaluated at the grid point $y_i$.
The positions of the grid points are given by the zeros of the Laguerre polynomial of order $N$ and are found by solving the implicit equation $L_N(y_i)=0$ for $y_i$. This procedure automatically divides the domain in elements of different size, with smaller elements near the surface of the bubble where larger gradients are expected. 
The resulting system of ODEs is solved using a fourth order Runge-Kutta implicit scheme with a variable time step size giving a maximum relative error of $10^{-9}$. We found that the choice $N=150$ yields numerically convergent results and guarantees that the far-field conditions are met at the last point $y_N$.
Numerical simulations performed with more refined grids, larger domains and smaller error threshold gave indistinguishable results.

\subsection{Validation of the code}
We validate the model and the numerical implementation by studying the dynamics of a bubble in the linear regime. Under the assumption of linearity, the material behaves as a Kelvin-Voigt viscoelastic solid. For small driving amplitudes the radial oscillations of the bubble can be expressed as $R(t) = R_0 \left[1 +x(t) \right]$, where $x(t)$ follows the same dynamics as the forcing, $x(t) = (\Delta R/R_0) \sin{\left(\omega \, t +\phi\right)}$, with a phase shift $\phi$. In this regime, Eqs. \eqref{genRP_nd}-\eqref{nondimens_eqs} reduce to a damped harmonic oscillator driven by the external pressure \cite{prosperetti1977thermal}:
\begin{equation}\label{harmonic_osc}
\ddot{x} + 2 \beta \dot{x} + \omega_0^2 \, x = \frac{\Delta p}{\rho R_0^2}\sin{\left(\omega \, t\right)} \, \, \, .
\end{equation} 
The damping coefficient, $\beta $, and the resonance frequency, $\omega_0$, are given by \cite{hamaguchi2015linear,jamburidze2017high}:
\begin{equation}\label{damp_resfreq_def}
\beta = \frac{2 \eta_s}{\rho R_0^2} \; \; ,  \;\; \omega_0^2= \frac{3 p_0+ 4\gamma/R_0 +4G}{\rho R_0^2} \;\;.
\end{equation}
In the linear regime the medium behaves as a Kelvin-Voigt material thus the yield-stress does not enter in the expression for the damping coefficient and the resonance frequency.

The solution of Eq. \eqref{harmonic_osc} gives the amplitude of oscillation, $\Delta R$, and the phase, $\phi$, as a function of the frequency:
\begin{equation}\label{linear_amplitude}
\frac{\Delta R}{R_0} = \frac{\Delta p/ \left(\rho R_0^2\right)}{\sqrt{\left(\omega_0^2-\omega^2\right)^2+4\beta^2\omega^2}}  \;\;\; , \;\;\; \phi = \arctan{\left(\frac{2\omega \beta}{\omega^2 - \omega_0^2} \right)} \;\;.
\end{equation}

\begin{figure}[h!]
\centering
\includegraphics[width=1.0\textwidth]{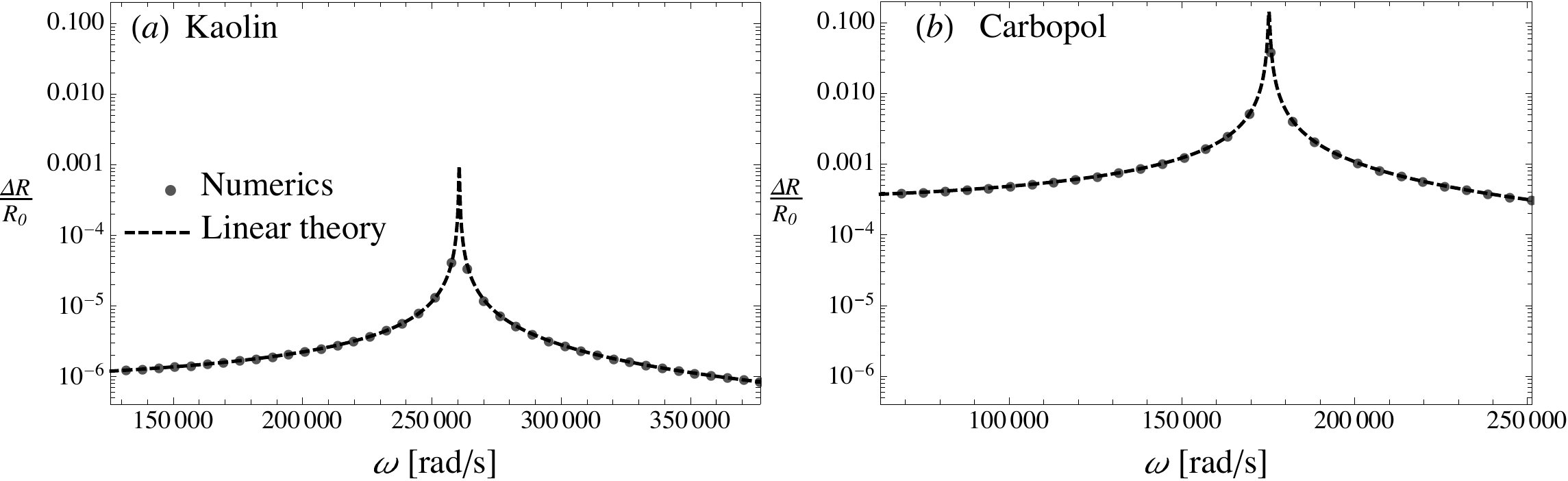}
\caption{Amplitude of the radial oscillations predicted by linear theory given by Eq. \eqref{linear_amplitude} and by the numerical solution of Eqs. \eqref{genRP_nd_Lag}-\eqref{nondimens_eqs_Lag} for a bubble with equilibrium radius $R_0=100 \, \rm{\mu m}$ driven by an acoustic field at (a) $\Delta p = 1 \, \rm{Pa}$ in a Kaolin suspension and (b) $\Delta p = 100 \, \rm{Pa}$ in a Carbopol gel.}
\label{fig1}
\end{figure}

We test the numerical solution of Eqs. \eqref{genRP_nd_Lag}-\eqref{nondimens_eqs_Lag} against the predictions of Eq. \eqref{linear_amplitude}. To avoid transient effects, we run simulations for $2000$ periods and we compute $\Delta R$ as the maximum radial excursion over the last period. In Figure \ref{fig1} we compare the amplitude of the radial oscillations predicted by the linear theory given by Eq. \eqref{linear_amplitude} with that obtained from the numerical solution. We considered the case of a bubble with equilibrium radius $R_0=100 \, \rm{\mu m}$, which is of the size used in the experiments performed by Jamburidze et al. \cite{jamburidze2017high}, driven at $\Delta p = 100 \, \rm{Pa}$ in the Carbopol gel (Figure \ref{fig1}(b)) and at $\Delta p = 1 \, \rm{Pa}$ in the Kaolin suspension (Figure \ref{fig1}(a)). Such a small acoustic pressure is required to ensure that the dynamics remains in the linear regime. One might question the need for such a smaller $\Delta p$ in Kaolin as opposed to Carbopol resulting in $\Delta R/R$ less than $10^{-3}$ in order to remain in the linear regime. The reason is that nonlinearity in this material is not induced by an increased amplitude of the radial oscillations, but by its yielding at smaller radial deformations caused by its larger elastic modulus, see related discussion in section 4. In Figure \ref{fig1}, the frequency at which the bubble experiences the largest radial excursion is approximately given by the resonance frequency $\omega_0$ because the damping coefficient is small. The numerical and analytical solutions show excellent agreement for all the frequencies investigated, thus showing that the numerical implementation of the EVP model correctly reduces to the a Kelvin-Voigt model for small deformations.

\section{Results and discussion}
\subsection{Analysis of bubble dynamics at resonance}

As a result of the harmonic change in pressure due to the acoustic driving, the bubble undergoes periodic compression and expansion. The dynamics of the bubble radius is strongly dependent on the amplitude of the pressure applied by the acoustic field and for sufficiently small $\Delta p$ it is given by the same harmonic function of the forcing. 

\begin{figure}[h!]
\centering
\includegraphics[width=1.0\textwidth]{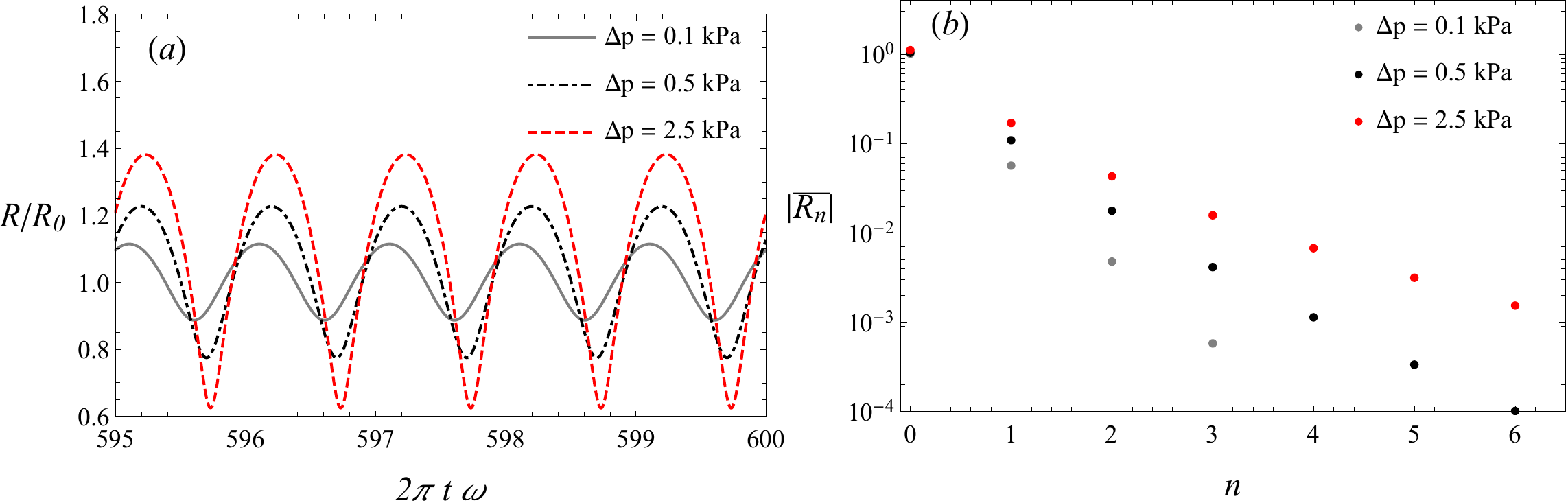}
\caption{(a) Numerical simulations of a $100 \, \rm{\mu m}$ bubble driven at $\omega = \omega_0$ in a Carbopol gel; (b) relative amplitude of the Fourier modes, $\left|\overline{R_n}\right|$, of $R(t)$ plotted as a function of the mode number $n$.}
\label{fig2}
\end{figure}

To explore the transition to the nonlinear regime, we report in Figure \ref{fig2}(a) the dynamics of a $100 \, \rm{\mu m}$ bubble driven at its resonance frequency, $\omega = \omega_0$, in the Carbopol gel. We discard the initial transient response of the bubble and we focus on its dynamics sufficiently far from the initial condition. As the acoustic pressure is increased, the amplitude of the radial excursion becomes larger, and the dynamics deviate from the single harmonic response expected in the linear regime. At the largest driving pressure, the bubble spends more time in its expanded than its contracted state because (i) in the former state the increased liquid inertia decreases its acceleration \cite{magnaudet2000motion} and (ii) the bubble pressure varies as $P_g \sim R^{-3}$, which changes much faster when the bubble radius is minimized. Both observations lead to an effect equivalent to the added mass effect in a translating bubble. To obtain a quantitative insight on the transition from the linear to the nonlinear regime, we plot in Figure \ref{fig2}(b) the amplitude of the Fourier modes $\left|\overline{R_n}\right|$ of $R(t)/R_0$. Figure \ref{fig2}(b) shows that the first harmonic is the mode with the largest amplitude for all driving investigated. In the case of $\Delta p = 0.1\rm{kPa}$ the amplitude of the modes higher than $1$ decay very fast with the mode number $n$. Conversely, in the case of $\Delta p = 2.5 \, \rm{kPa}$ the amplitude of the high-order modes decays slower with $n$ and multiple harmonics play a role in the radial response. The significant coupling between different modes $\left|\overline{R_n}\right|$ is a signature of the nonlinear dynamics of the bubble at large driving pressures. 

\begin{figure}[h!]
\centering
\includegraphics[width=1.0\textwidth]{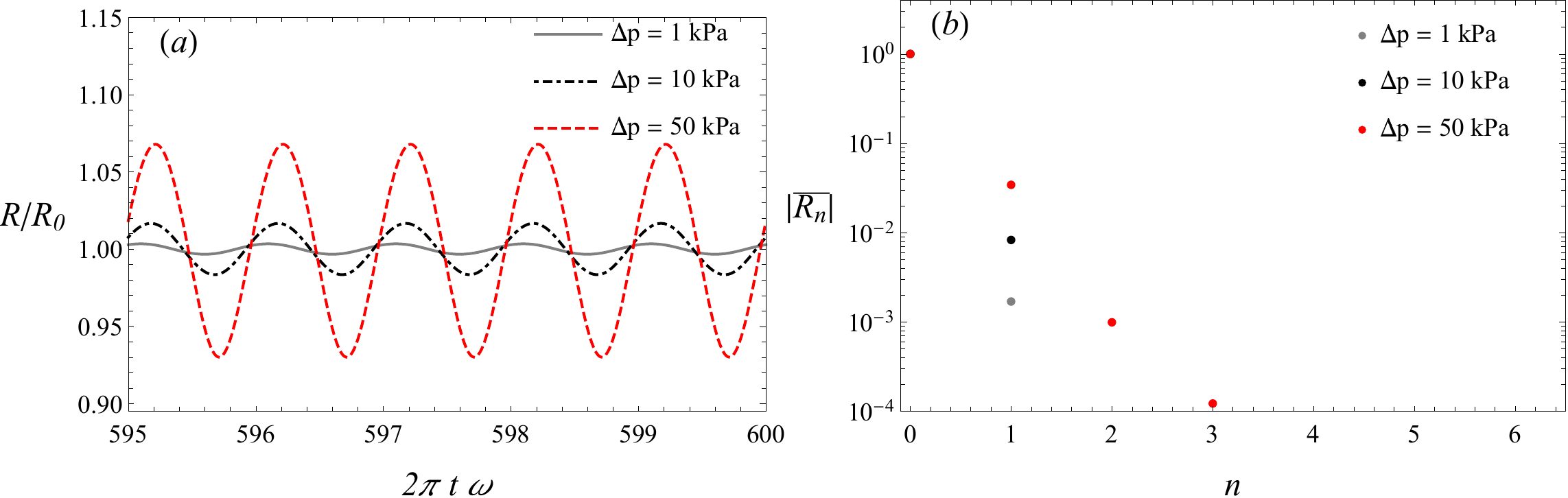}
\caption{(a) Numerical simulations of a $100 \rm{\mu m}$ bubble driven at $\omega = \omega_0$ in a Kaolin suspension; (b) amplitude of the Fourier modes, $\left|\overline{R_n}\right|$, of $R(t)$ plotted as a function of the mode number $n$.}
\label{fig2kaolin}
\end{figure}
The radial dynamics of a $100 \, \rm{\mu m}$ bubble driven at its resonance frequency in the Kaolin suspension is reported in Figure \ref{fig2kaolin}(a). In contrast to the case of a bubble driven in the Carbopol gel, Figure \ref{fig2kaolin}(a) suggests that the radial dynamics is given by a single harmonic for all the driving amplitudes. This is confirmed by Figure \ref{fig2kaolin}(b), which shows that the amplitude of the modes $n>1$ is significantly smaller than the mode $n=1$ for all the $\Delta p$ investigated. Nevertheless the bubble dynamics are nonlinear as can be seen in the change of mode amplitude as a function of the driving amplitude. In the linear regime one expects the first mode to be linearly proportional to $\Delta p$. Figure \ref{fig2kaolin}(b) shows that increasing $\Delta p$ by ten times, from $1\rm{kPa}$ to  $10\rm{kPa}$, results in a fivefold increase of $\left|\overline{R_1}\right|$, which is a signature of the nonlinear dynamics of the bubble, although this signature is weaker than in Carbopol. In other words, in Kaolin, even when much higher pressure amplitudes are used than in Carbopol, nonlinearity cannot be detected either by the amplitude of the radial oscillations or by the (quite small) Fourier modes that are hardly present.

\subsection{Conditions for oscillation-induced yielding}
As the driving pressure is increased the large-amplitude oscillations of the bubble generate considerable elastic stresses, potentially yielding the surrounding material. The periodic expansion and compression of the bubble generates extensional and compressive strains in the yield-stress fluid. If the amplitude of the acoustic pressure, $\Delta p$, is sufficiently large, the periodic elongational stresses due to the radial oscillations of the bubble can be larger than the yield stress. For a fixed set of constitutive parameters and a given bubble equilibrium size, there exists a frequency-dependent critical driving pressure, $\Delta p_{\text{crit}}$, above which the material around the bubble yields during an oscillation cycle. The maximum normal stress difference, $\tau_{rr}-\tau_{\theta \theta}$, occurs at the bubble surface and decays to zero at infinity. Thus, $\Delta p_{\text{crit}}$ is defined as the minimum pressure amplitude, $\Delta p$, for which the Von Mises criterion,
\begin{equation}\label{yielding_cond}
\left\vert \tau_{rr}(R)-\tau_{\theta \theta}(R) \right\vert = \sqrt{3} \tau_y\, \, ,
\end{equation} 
is satisfied at the bubble surface at least at one instant during a cycle.
To find $\Delta p_{\text{crit}}$ one has to solve the system of Eqs. \eqref{genRP_nd_Lag}-\eqref{nondimens_eqs_Lag} numerically for different acoustic pressure amplitudes and frequencies and find the minimum $\Delta p$ for which Eq. \eqref{yielding_cond} is satisfied.
An estimate of $\Delta p_{\text{crit}}$ can be obtained by assuming that the dynamics of the bubble and that of the yield-stress fluid are linear until yielding occurs. The validity of this assumption is verified later through numerical simulations. In the linear regime, it is $\tau_{\theta \theta} =-\tau_{rr}/2$ \cite{allen2000dynamics,jamburidze2017high} and the elastic stress is linearly related to the strain:
\begin{equation}\label{lin_el_stress}
\tau_{rr}=-4G (R^3-R_0^3)/(3r^3) \, \, .
\end{equation}
We evaluate Eq. \eqref{lin_el_stress} at $r=R$, with the assumption of small radial oscillations $R=R_0 \left[1+x(t)\right]$:
\begin{equation}
\tau_{rr}(R,t)=-4G x(t) \, \, .
\end{equation}
Since in the linear regime $x(t) = \Delta R/R_0 \sin{\left(\omega t +\phi \right)}$, the maximum amplitude during each cycle is given by $\Delta R/R_0$. To evaluate $\Delta R/R_0$ we use Eq. \eqref{linear_amplitude}, resulting in the following maximum value of the radial stress at the bubble surface during each cycle:
\begin{equation}\label{maxradstress}
\tau_{rr, \text{max}}(R)=-4G \frac{\Delta p_{\text{crit}}/ \left(\rho R_0^2\right)}{\sqrt{\left(\omega_0^2-\omega^2\right)^2+4\beta^2\omega^2}} \, \, ,
\end{equation}
with $\beta$ and $\omega_0$ defined in Eq. \eqref{damp_resfreq_def}. By inserting the maximum radial stresses given by Eq. \eqref{maxradstress} in the yielding criterion given by Eq. \eqref{yielding_cond} and considering that $\tau_{\theta\theta}=-\tau_{rr}/2$, we obtain an equation for the critical driving pressure for yielding:
\begin{equation}\label{crit_driv_press}
\Delta p_{\text{crit}} = \frac{\rho R_0^2 \tau_y}{2 \sqrt{3} G} \sqrt{\left(\omega_0^2-\omega^2\right)^2+4\beta^2\omega^2} \, \, \, .
\end{equation}
Eq. \eqref{crit_driv_press} shows that $\Delta p_{\text{crit}}$ depends linearly on the yield stress of the material and varies strongly with the driving frequency. The critical yielding pressure amplitude has a minimum at the resonance frequency of the bubble, $\omega_0$. 
If yielding of the material due to the dynamics of the bubble is desired, e.g. to promote release of bubbles of a known size from a yield-stress fluid, Eq. \eqref{crit_driv_press} can be used as a rule of thumb to select the acoustic pressure and its frequency. 

The dynamics of the bubble can be nonlinear even before yielding, due to the nonlinear inertial and elastic terms in Eqs. \eqref{genRP_nd_Lag1} and \eqref{nondimens_eqs_Lag}. It follows that the assumptions used to derive Eq. \eqref{crit_driv_press} might break down. 
To verify its relevance, we compare the linear yielding criterion $\Delta p_{\rm crit}$ obtained from Eq. \eqref{crit_driv_press} with the numerical results obtained from the solution of the non-linear Eqs. \eqref{genRP_nd_Lag}-\eqref{nondimens_eqs_Lag}.
We run the simulations for $2000$ driving periods, always starting from the rest state, and discard the initial transient response. We consider the material yielded if at any instant during the last ten cycles Eq. \eqref{yielding_cond} is satisfied.
For a fixed frequency we run simulations at increasing $\Delta p$ until yielding is detected at a single instant during a cycle. This value of $\Delta p$ is considered the $\Delta p_{\text{crit}}$ for that particular frequency. By repeating this process for different frequencies we construct the curve $\Delta p_{\text{crit}}(\omega)$.

\begin{figure}[h!]
\centering
\includegraphics[width=1.0\textwidth]{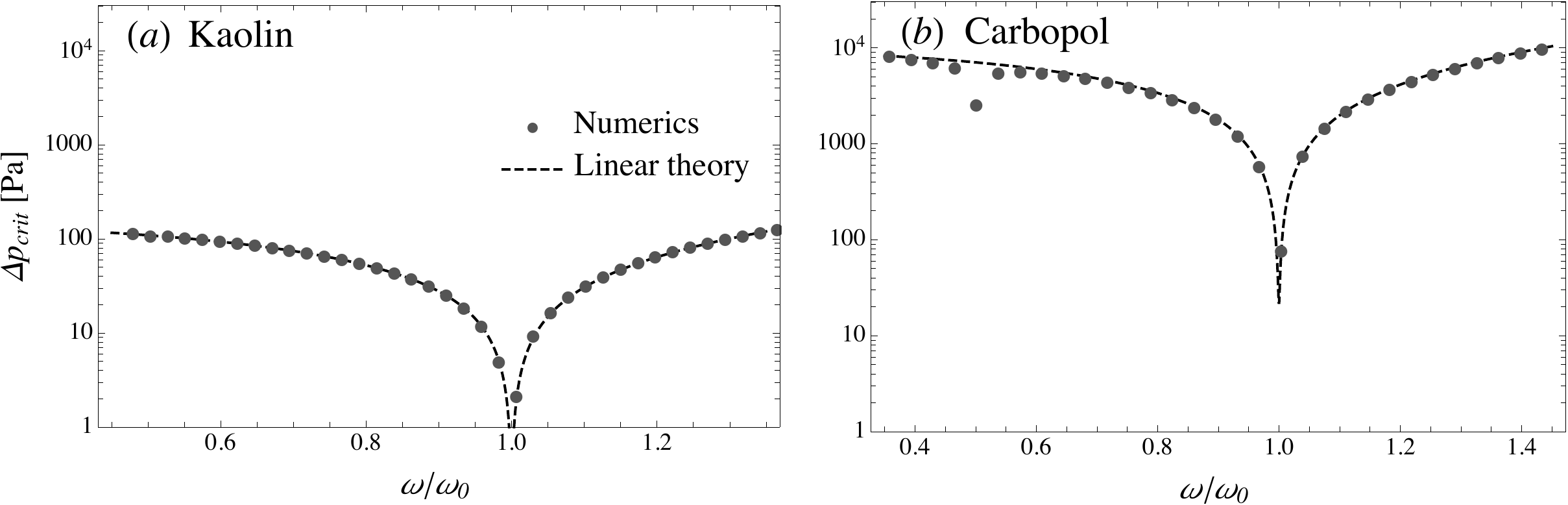}
\caption{Critical pressure amplitude for which yielding of the medium is triggered by the bubble oscillations as a function of the angular frequency. (a) critical pressure for a bubble in the Kaolin suspension and (b) critical pressure in the case of the Carbopol gel. The dashed line is the prediction of the linear theory given by Eq. \eqref{linear_amplitude}, the solid symbols represent the critical pressure obtained by the numerical solution of Eqs. \eqref{genRP_nd_Lag}-\eqref{nondimens_eqs_Lag}. The bubble size considered in this figure is $R_0=100 \, \rm{\mu m}$.}
\label{fig3}
\end{figure}

In Figure \ref{fig3} we plot the critical pressure amplitude as a function of the angular frequency as predicted by Eq. \eqref{crit_driv_press} and by the numerical simulations. We considered the case of a bubble with an equilibrium radius $R_0 = 100 \, \rm{\mu m}$ suspended in the two  yield-stress fluids considered, which have very different elastic moduli. In Figure \ref{fig3}(a) we show the results obtained with the stiff Kaolin suspension and in Figure \ref{fig3}(b) we consider the rheological parameters for the soft Carbopol gel. 

Figure \ref{fig3} shows that $\Delta p_{\text{crit}}$ depends strongly on the driving frequency and shows a pronounced minimum at $\omega = \omega_0$ that is the resonance frequency of the bubble. At resonance, the driving pressure that yields the medium can be orders of magnitude smaller than that obtained off resonance. At frequencies that are much smaller than the resonance frequency of the bubble, $\Delta p_{\text{crit}}$ approaches a constant value given by $\Delta p_{\text{crit}}\approx \tau_y (4G+3 p_0 +4 \gamma/R_0)/(2\sqrt{3} G)$. This expression and Eq. \eqref{crit_driv_press} indicate that the yielding pressure amplitude is inversely proportional to the elastic modulus. At frequencies larger than the resonance frequency, $\Delta p_{\text{crit}}$ is a linearly increasing function of $\omega$. It is noteworthy that a stiffer EVP material generally requires smaller $\Delta p$ to yield, i.e. it yields easier. In the case of a stiff yield-stress fluid, Figure \ref{fig3}(a) shows that the linear theory gives a very good prediction of the $\Delta p_{\text{crit}}$ for all the angular frequencies explored and that very low pressure amplitudes are required to yield the material. The good agreement between the linear theory and the simulations is a consequence of the small strains $R/R_0 \approx \tau_y/G$ at which a very stiff material yields. Since for the stiff Kaolin suspension it is $\tau_y/G=0.00045$, prior to yielding the oscillations of the bubble are very small and the dynamics is in the linear regime.

\begin{figure}[h!]
\centering
\includegraphics[width=1.0\textwidth]{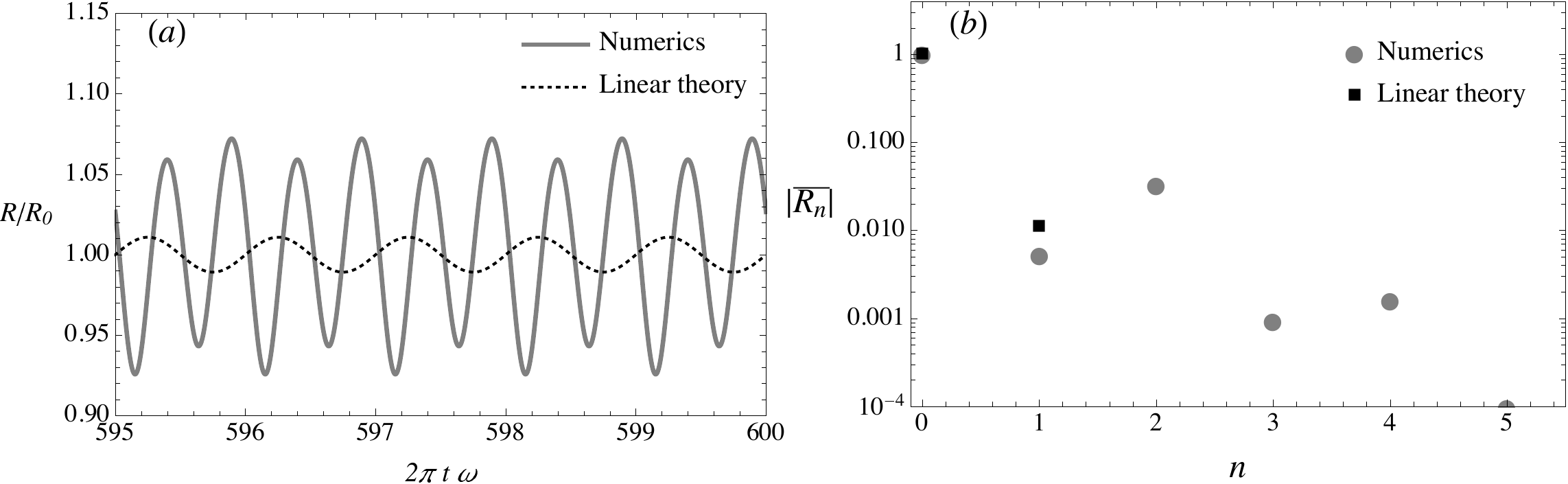}
\caption{(a) Response of a $100 \, \rm{\mu m}$ bubble driven at $\omega = \omega_0/2$ and $\Delta p = 2.5 \,  \rm{kPa}$ in a Carbopol gel. (b) Amplitude of the Fourier modes $\left|\overline{R_n} \right|$ of the radial dynamics. Linear theory predicts a single harmonics response $n=1$. Numerical simulations reveal that the dominant mode is $\left|\overline{R_2} \right|$, which corresponds to oscillations at twice the driving frequency. The total radial excursion is larger than that predicted by the linear theory.}
\label{fig4}
\end{figure}

In the case of the soft Carbopol gel, shown in Figure \ref{fig3}(b), the critical pressure computed with the numerical simulations is in good agreement with the linear approximation for almost all the frequencies, except for $\omega \approx \omega_0/2$. The discrepancy observed at $\omega \approx \omega_0/2$ is a signature of the weakly nonlinear response of the bubble that is neglected in the linear theory. The characteristic yield strain $\tau_y/G = 0.1$ of the Carbopol gel is much larger than that of the Kaolin suspension and at this strain the dynamics of the bubble can be weakly nonlinear. In Figure \ref{fig4} we show that for $\omega = \omega_0/2$, the radial oscillations depart from a single harmonic response predicted in the linear regime and the bubble experiences multiple harmonics with the dominant one being twice the angular frequency of the acoustic driving  $\omega$. This is confirmed by investigating the amplitude of the Fourier modes reported in Figure \ref{fig4}(b), which show that the largest mode $\left|\overline{R_n} \right|$ is given by $n=2$. The additional harmonics shown in Figure \ref{fig4}(b) induce larger radial excursions compared to those predicted by the linear theory, hence resulting in larger strains and a smaller $\Delta p_{\text{crit}}$. 
In summary, the results obtained with Carbopol and Kaolin suggest that Eq. \eqref{crit_driv_press} is a very good estimate of the critical pressure for materials with characteristic yield strain, $\tau_y/G$ smaller than one. 

\subsection{Dynamics of the yield surface}
In the case of a driving pressure larger than $\Delta p_{\text{crit}}$, part of the material surrounding the bubble is yielded and behaves as a liquid and the remaining part behaves as a solid. As a result, the bubble oscillates in a cavity with a time-dependent radius, filled by a viscoelastic liquid and surrounded by an elastic solid. This situation has been studied by Vincent et al. \cite{vincent2012birth, vincent2014fast} in the context of cavitation in trees. The liquid and the solid regions are separated by the yield surface whose instantaneous position, $r_y(t)$, is defined as the radial coordinate at which the Von Mises criterion is satisfied: $\left \vert \tau_{rr}\left(r_y(t),t\right)-\tau_{\theta \theta}\left(r_y(t),t\right)\right\vert =\sqrt{3} \tau_y$. The periodic compression and expansion of the bubble generate cyclic elongational stresses that result in a time-dependent yield surface $r_y(t)$. Since a bubble trapped in a yield-stress fluid can only rise when the surrounding material is yielded, it is interesting to investigate the evolution of the yield surface as its dynamics could have a strong impact on the rising velocity of the bubble. 
\begin{figure}[h!]
\centering
\includegraphics[width=1.0\textwidth]{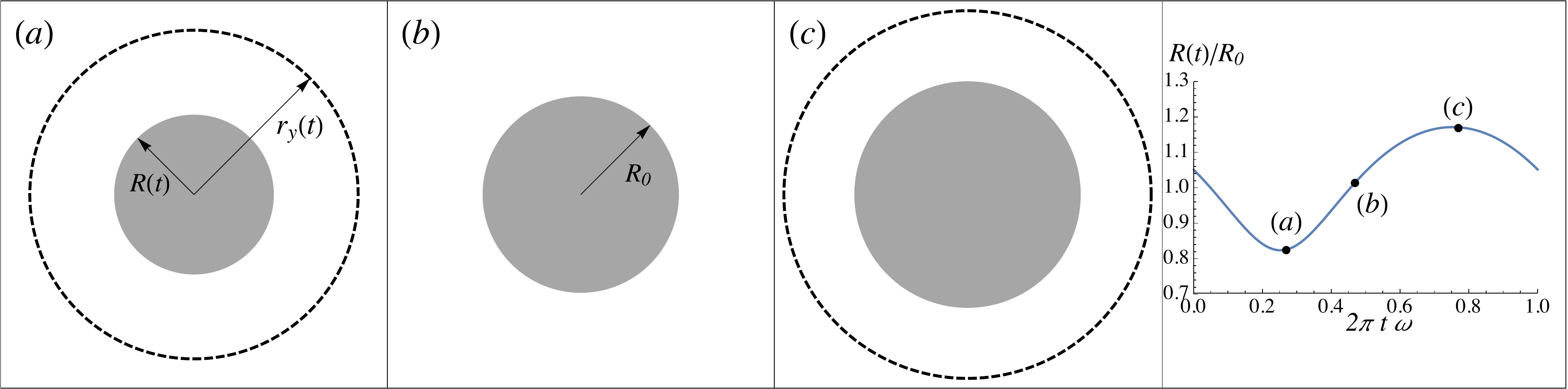}
\caption{Snapshots of the bubble dynamics and of the yielded region computed through numerical simulations at three different instants within one period: a) minimum radius, b) equilibrium radius, c) maximum expansion. The grey area represents the bubble, the dashed line represents the yield surface. The rightmost panel shows the dynamics of the bubble radius. The dimensionless numbers considered in this figure are: $De = 7.36$, $Bn= 0.014 $, $Re = 411$, $\Delta P = 16.6$ and $P_\text{stat} = 72.37$ , corresponding to a $R_0 = 100 \, \rm{\mu m}$ bubble in Carbopol.}
\label{fig5}
\end{figure}

\begin{figure}[h!]
\centering
\includegraphics[width=1.0\textwidth]{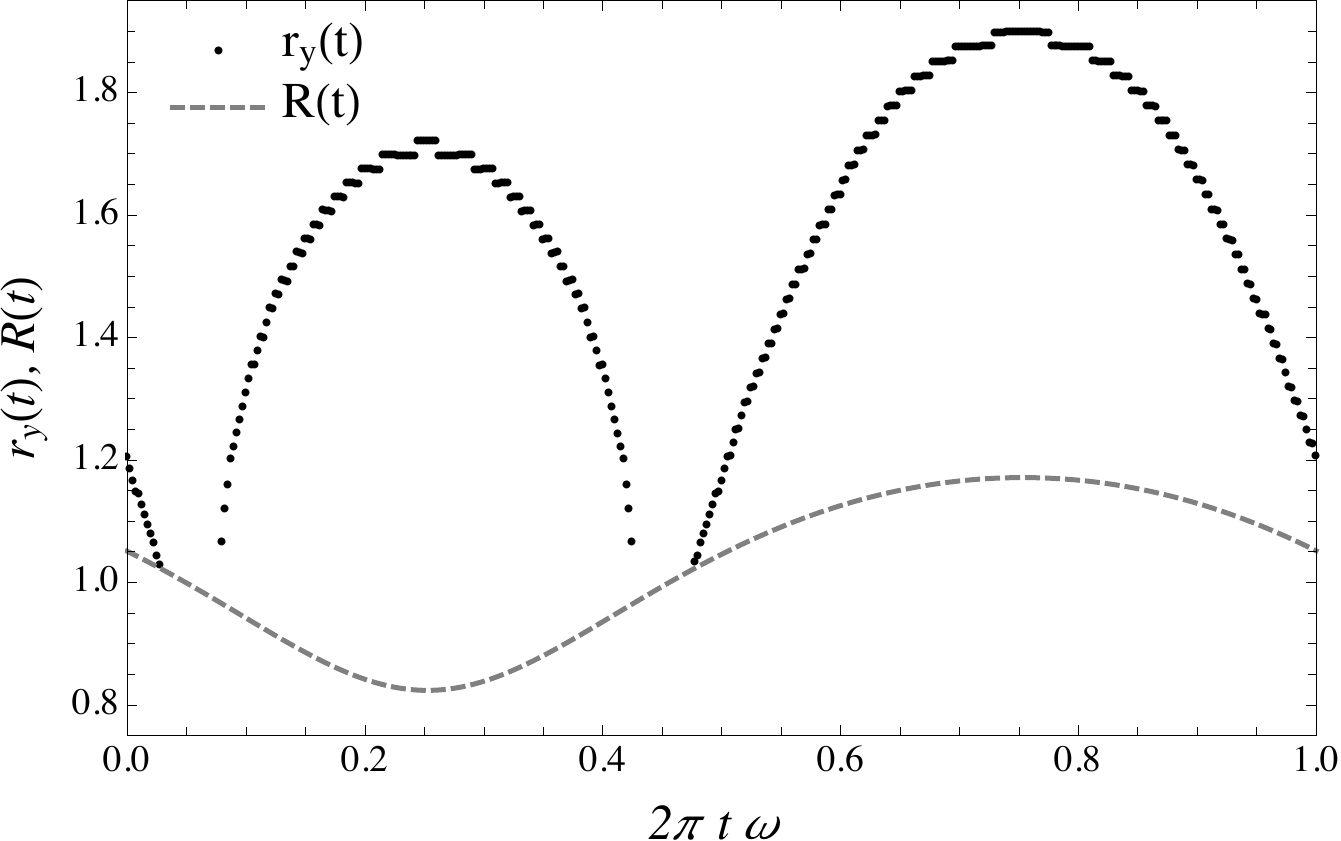}
\caption{Numerical simulations of the dynamics of the yield surface $r_y(t)$ and of the bubble radius $R(t)$ during one cycle for the same parameters used in Figure \ref{fig3}. The material unyields and yields twice per cycle when the bubble radius is close to its equilibrium radius and the elastic strains are small.}
\label{fig6}
\end{figure}

In Figure \ref{fig5} we show snapshots of the bubble radius and of the position of the yield surface at three different instants during one cycle: minimum radius, equilibrium radius and maximum expansion. We consider the case of a bubble of equilibrium radius $R_0 =100 \, \rm{\mu m}$ trapped in a Carbopol gel and driven at $\omega = 0.9 \, \omega_0 $ and $\Delta p=10 \, \rm{kPa}$. To avoid transient effects, the period shown in Figure \ref{fig5} is chosen sufficiently far away from $t=0$. In Figure \ref{fig5}(a) the bubble is compressed, $R < R_0$, and the Carbopol gel surrounding the bubble is yielded. As the bubble radius increases to reach its equilibrium value, the strain and the elastic stresses decrease thus the yield surface moves towards the surface of the bubble. Eventually, the Carbopol unyields for $R = R_0$, see Figure \ref{fig5}(b). During the subsequent expansion of the bubble, the strain and the elastic stresses increase again, the Carbopol gel yields and the yielded region grows, see Figure \ref{fig5}(c).  
To get a more detailed insight of the dynamics of the yield surface, we plot $r_y(t)$ and the bubble radius $R(t)$ in Figure \ref{fig6} during one period. When the material is unyielded the yield surface is not defined, which explains why $r_y(t)$ in Figure \ref{fig6} is clipped for certain time intervals. The Carbopol gel unyields and then yields twice per cycle during the compression and the expansion phases as $R$ goes through $R_0$. This is a consequence of the change of sign of the normal stress difference between the compression and expansion phases, which implies that the normal stress difference must go through zero. It follows that, for $R \approx R_0$, the deviatoric part of the stress tensor is smaller than the yield stress and the material unyields everywhere.

\begin{figure}[h!]
\centering
\includegraphics[width=1.0\textwidth]{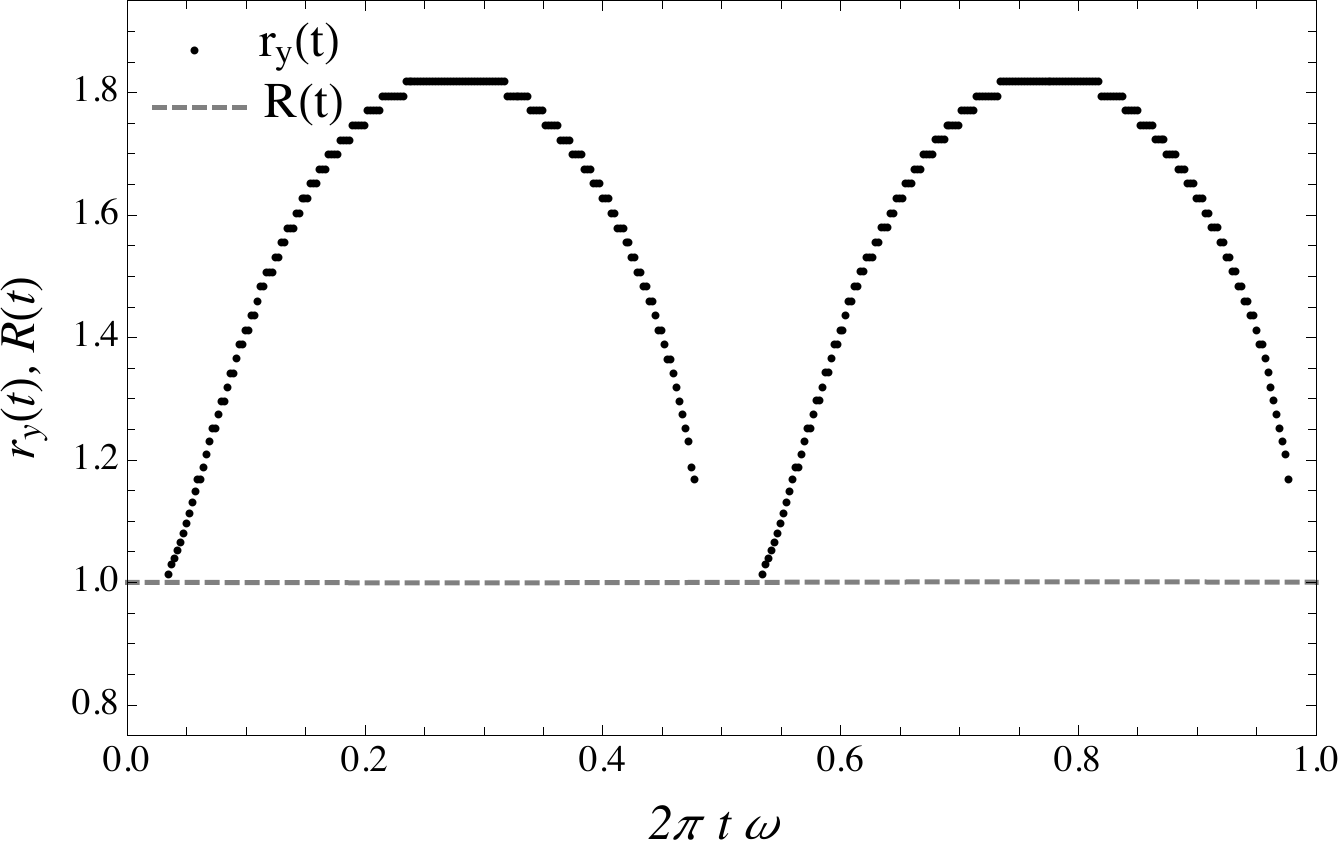}
\caption{Numerical simulations of the dynamics of the yield surface $r_y(t)$ and of the bubble radius $R(t)$ during one cycle, for a bubble oscillating in the Kaolin suspension. The dimensionless numbers considered in this figure are: $De = 0.042$, $Bn= 0.011 $, $Re = 105$, $\Delta P = 0.02$ and $P_\text{stat} = 72.37$.}
\label{fig7}
\end{figure}

The dynamics of the yield surface and of the bubble radius in a Kaolin suspension are shown in Figure \ref{fig7}. We consider the case of a bubble of equilibrium radius $R_0 = 100 \, \rm{\mu m}$ driven at $\omega = 0.9 \, \omega_0 $ and $\Delta p= 0.16 \,  \rm{kPa}$, which corresponds to $5.5$ times the critical pressure.
Due to the large elastic modulus of the Kaolin suspension the elastic stresses are sufficiently large to yield part of the material, despite the very small oscillations of the bubble. 
In contrast to the case of a bubble oscillating in the Carbopol gel, the yielded region in the Kaolin suspension behaves essentially as a viscous fluid because $De$ is small. 
Despite this difference, Figure \ref{fig7} reveals that $r_y(t)$ obtained in the case of a Kaolin suspension is qualitatively similar to that obtained for a Carbopol gel.  

We derive an approximate expression for the dynamics of the yield surface under the assumptions that the dynamics of the bubble is linear and that the material behaves as a Kelvin-Voigt solid everywhere. These assumptions are reasonable if the characteristic yield strain is small $\tau_y/G \ll 1$, if the material behaves mostly elastically in the yielded region $De \gg 1$, and if the driving pressure is close to the critical pressure $\Delta p \approx \Delta p_{\text{crit}}$. The stresses in the material are then given by:
\begin{equation}\label{stressofr}
\tau_{rr}(r,t)=-2\tau_{\theta \theta}(r,t)=-4 G \frac{R_0^2 }{r^3} \frac{\Delta R}{R_0} \sin{\left(\omega t+ \phi\right)} \,\, \, .
\end{equation}
The position of the yield surface is given by the radial coordinate at which the Von Mises yielding criterion is satisfied, giving the implicit equation:
\begin{equation}\label{yieldsurftt0}
\left| \tau_{rr}(r_y(t),t) - \tau_{\theta \theta}(r_y(t),t) \right| = \sqrt{3} \tau_y \,\,\, ,
\end{equation}
subjected to the constraint that $r_y(t)>R_0$. If at any instant it is $r_y(t) \leq R_0$, the material is unyielded.
Substitution of the stresses given by Eq. \eqref{stressofr} into Eq. \eqref{yieldsurftt0} gives an equation for $r_y(t)$:
\begin{equation}\label{yieldsurflin}
r_y(t) =  R_0 \sqrt[\leftroot{-1}\uproot{2}\scriptstyle 3]{\frac{2 \sqrt{3} G \Delta R}{\tau_y R_0} \, \left|\sin{\left(\omega t+ \phi\right)}\right|} \, \, \, ,
\end{equation}
with $\Delta R/R_0$ and $\phi$ given by Eq. \eqref{linear_amplitude}.
\begin{figure}[h!]
\centering
\includegraphics[width=1.0\textwidth]{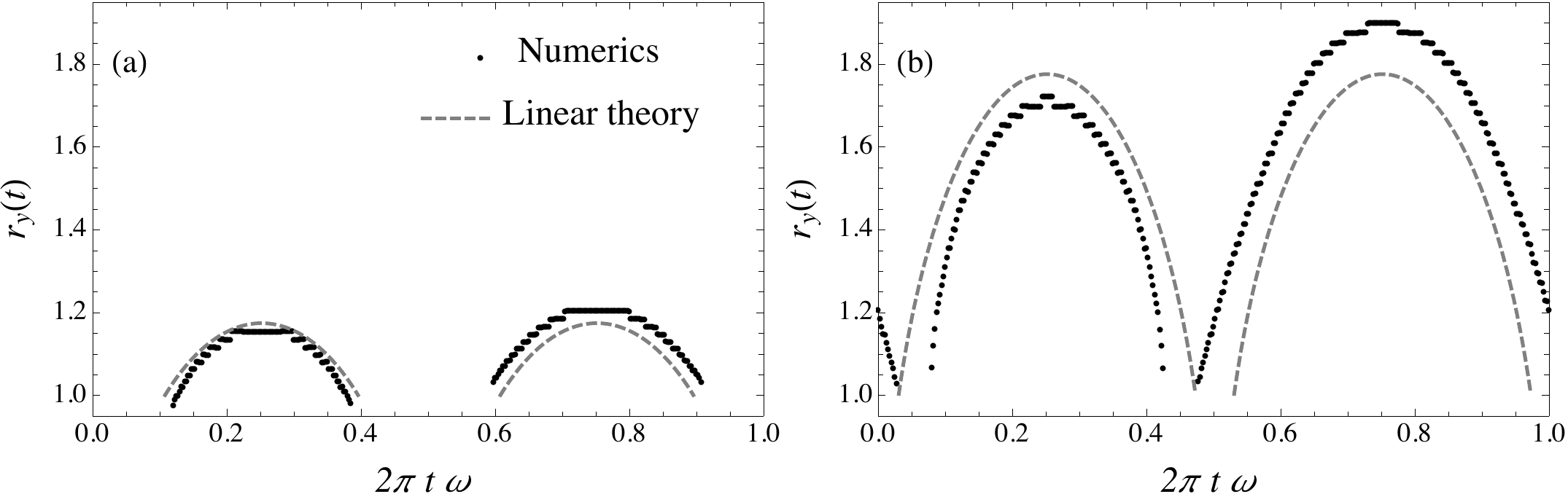}
\caption{Comparison of the dynamics of the yield surface in the Carbopol gel as predicted by the linear approximation given by Eq. \eqref{yieldsurflin} (dashed line) and by numerical simulations (solid symbols). for $De = 7.36$, $Bn= 0.014 $, $Re = 411$, $P_\text{stat} = 72.4$ and for two different acoustic driving pressures: (a) $\Delta P = 5$ corresponding to $\Delta p = 1.62 \Delta p_{\text{crit}}$ and (b) $\Delta P = 16.6$ corresponding to dimensional driving amplitude $\Delta p = 5.6 \Delta p_{\text{crit}}$. }
\label{fig8}
\end{figure}

Since Eq. \eqref{yieldsurflin} is derived under the assumption of linear bubble dynamics, we expect it to break down for driving pressures much larger than the critical pressure $\Delta p_{\text{crit}}$. 
We explore the range of validity of Eq. \eqref{yieldsurflin} by finding the position of the yield surface through numerical simulations at different $\Delta p/ \Delta p_{\text{crit}}$, with $\Delta p_{\text{crit}}$ computed from Eq. \eqref{crit_driv_press}. In Figure \ref{fig8}, we report the evolution of the yield surface predicted by the numerical simulations and by Eq. \eqref{yieldsurflin} for a bubble with equilibrium radius $R_0=100 \, \rm{\mu m}$ driven by an acoustic field at $\omega = 0.9 \omega_0$ in the Carbopol gel. The Deborah number corresponding to this case is $De =7.36$. In figure \ref{fig8}(a) the bubble is driven at a pressure close to the critical pressure, $\Delta p = 1.62  \Delta p_{\text{crit}}$, and the dynamics of the yield surface given by the linear approximation given by Eq. \eqref{yieldsurflin} is very close to that obtained in the numerical simulation. As expected, Figure \ref{fig8}(b) shows that, by increasing the acoustic driving to $\Delta p = 5.6 \Delta p_{\text{crit}}$, the dynamics becomes pronouncedly nonlinear and the linear theory fails to predict the evolution of the yield surface quantitatively. The linear theory systematically overpredicts $r_y(t)$ in the first half of the cycle and underpredicts it in the second half. This is a consequence of the linearization, which neglects the advection of the yield surface due to the displacement of the bubble surface. 

\subsection{Impact of yielding on the radial dynamics}
In this section we explore the impact of yielding on the radial dynamics of a bubble. To highlight the effects of visco-plastic deformations we compare the dynamics of a $100 \rm{\mu m}$ bubble in an EVP fluid and in a Neo-Hookean solid with the same elastic modulus. If any difference between the two behaviors is observed, it must be due to the yielding of the medium.
\begin{figure}[h!]
\centering
\includegraphics[width=1.0\textwidth]{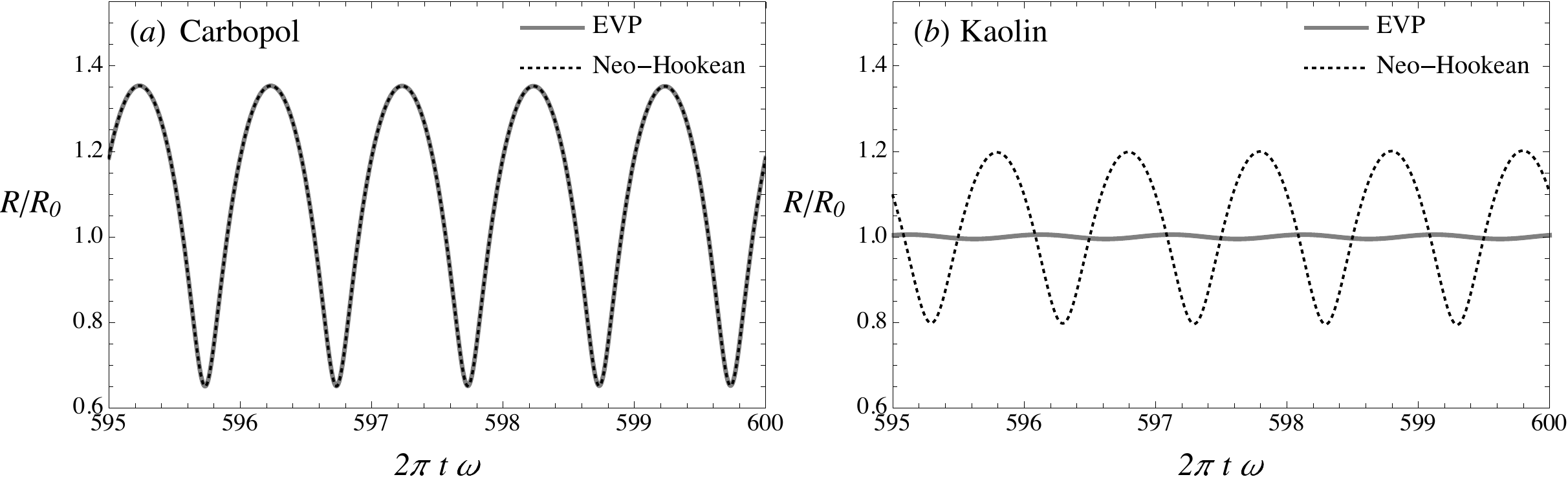}
\caption{Dynamics of a $100 \, \rm{\mu m}$ bubble driven at $\omega = \omega_0$ and $\Delta p = 2 \, \rm{kPa}$ in (a) Carbopol gel and (b) Kaolin suspension, compared to its dynamics in a Neo-Hookean elastic solid with same elastic modulus.}
\label{fig9}
\end{figure}
In Figure \ref{fig9} we plot the dynamics of a bubble driven at $\Delta p = 2 \, \rm{kPa}$ and at $\omega=\omega_0$ in the Carbopol gel and in the Kaolin suspension, compared to its dynamics in a Neo-Hookean solid. In the case of the Carbopol gel, Figure \ref{fig9}(a) shows that the dynamics of the bubble is indistinguishable from that predicted in a Neo-Hookean solid. Due to the small elastic modulus of the Carbopol gel compared to the Kaolin suspension, the relaxation time of the liquid in the fluidized region is much larger than the driving frequency and $De = 7.73$. It follows that the yield-stress material behaves as an elastic solid both in the yielded and in the unyielded region, thus making the dynamics of the bubble identical to that predicted by the Neo-Hookean model. Conversely, the dynamics of a bubble oscillating in the Kaolin suspension is markedly different from that predicted by a Neo-Hookean model, with the oscillations being significantly damped. In this case, it is $De=0.018$ and the yielded region behaves as a viscous fluid. It follows that yielding of the material manifests itself as a larger damping compared to that expected for a Neo-Hookean solid.
These findings suggest that, in the case $De <1$, it should be possible to experimentally verify if the material has yielded by observing the dynamics of the bubble. Ideally, if by increasing the acoustic pressure above $\Delta p_\text{crit}$, a qualitative change in the dynamics of the bubble is observed due to yielding of the material, one might be able to measure the yield-stress or at least identify yielding at high frequencies.

\begin{figure}[h!]
\centering
\includegraphics[width=1.0\textwidth]{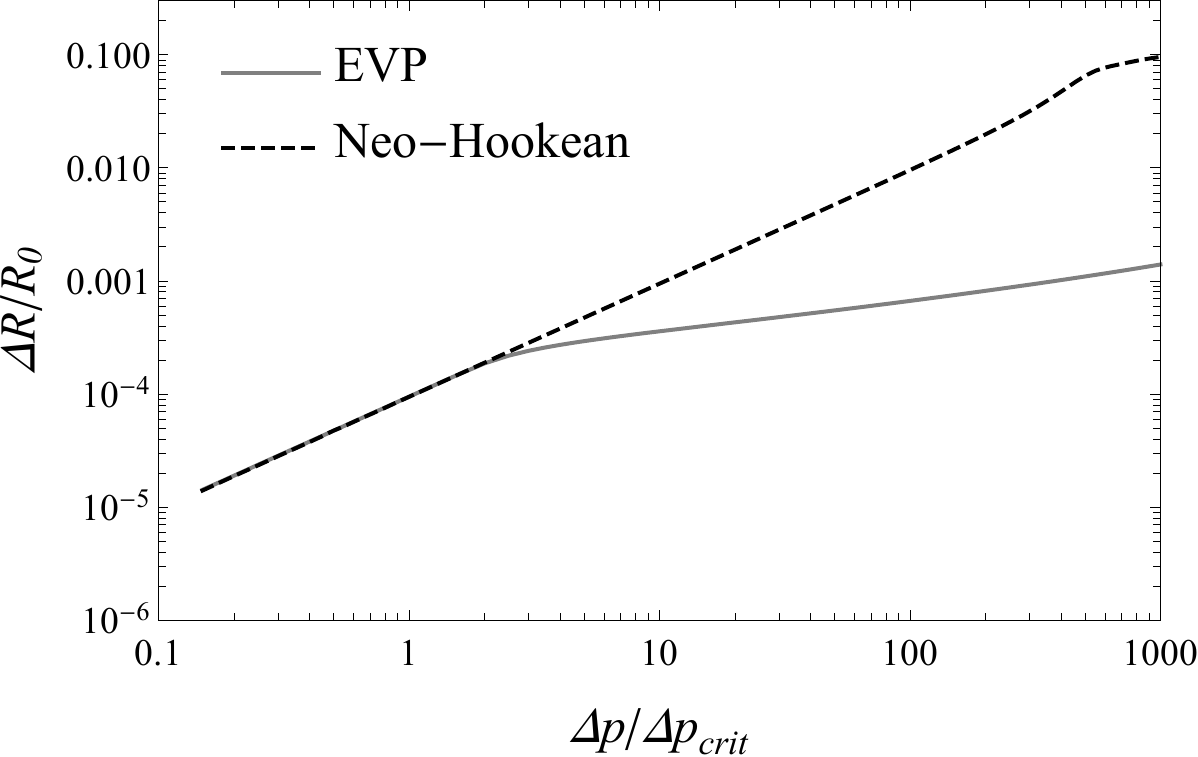}
\caption{Maximum radial excursion during one period of a $100 \,  \rm{\mu m}$ bubble driven at $\omega = \omega_0$ and different pressures. The rheological parameters are those for the Kaolin suspension. The comparison between the EVP model and the Neo-Hookean model highlights the extra damping experienced by the bubble due to yielding for $\Delta p > \Delta p_\text{crit}$.}
\label{fig10}
\end{figure}

We test this hypothesis by computing the maximum radial excursion of a bubble in a Kaolin suspension, for the same parameters used in Figure \ref{fig9} but varying the driving amplitude. The comparison with the Neo-Hookean model reported in Figure \ref{fig10} shows that the amplitude of oscillations is significantly lower than that predicted in an elastic solid for driving pressures larger than $\Delta p_\text{crit}$. This is a consequence of the additional dissipative processes taking place in the yielded region. The onset of the additional damping is not sharp at $\Delta p=\Delta p_\text{crit}$, because for pressure slightly larger than the critical pressure the size of the yielded region is small and the viscous stresses do not significantly impact the dynamics of the bubble. Finally,  Figure \ref{fig10} shows that for pressures $\Delta p > \Delta p_\text{crit}$ the oscillations of the bubble in the EVP fluid grow less than linearly with the driving amplitude. The sublinear increase of the oscillation amplitude is a consequence of the increase of viscous dissipation as the yielded region grows. The significant signature that yielding can have on the dynamics of the bubble suggests that a potential protocol for investigating yielding in experiments using acoustically-driven microbubbles is to increase the driving power at a fixed frequency progressively. These findings have implications for bubble removal: as the yielded region grows, most of the power input by the pressure waves is lost to viscous dissipation. As a consequence, there might be an optimal choice for the power that maximizes the efficiency of the bubble release process.

\section{Conclusions}
\label{sec:conclusions}
We have investigated the dynamics of a bubble driven by an oscillating pressure field in an incompressible and elastic yield-stress fluid using numerical simulations and an approximate linear theory. We modelled the rheological behavior of the fluid using a recently developed constitutive model \cite{saramito2009new} that takes into account both elastic and visco-plastic deformations. By assuming that the bubble remains spherical during the pressure driving, we reduced the problem to a set of integro-differential equations that we solve numerically using a Gauss-Laguerre method for the spatial integral and a fourth order implicit Runge-Kutta time integration method. To explore the effects of different rheological parameters, we considered the case of a bubble driven by an acoustic field in a soft Carbopol gel and in a stiff Kaolin suspension. 

For a given bubble there exists a frequency-dependent critical pressure at which the oscillations of the bubble yield the material. The critical pressure varies significantly with the frequency and it shows a pronounced minimum at the resonance frequency of the bubble. The critical pressure is very well approximated by an analytical formula derived under the assumption of linear bubble dynamics. In the case of an acoustic pressure larger than the critical pressure a dynamic yield surface is developed inside the yield-stress fluid in the immediate environment of the bubble. We found that the position of the yield surface evolves significantly during one period both in the Carbopol gel and Kaolin suspension. The material unyields and then subsequently yields twice per period as the bubble goes through its equilibrium configuration. This is a consequence of the small elastic stress imparted to the yield-stress fluid by a bubble that is close to its equilibrium configuration. We developed an equation for the dynamics of the yield surface based on a linear approximation of the bubble oscillations. The linear theory is in good agreement with the fully nonlinear numerical simulations for $\Delta p \approx \Delta p_{\text{crit}}$ but deviates for larger driving amplitudes for which the assumption of linearity breaks down.

Finally, we explored the impact of yielding of the medium on the radial oscillations of the bubble. In the case of soft yield-stress fluids with elastic modulus in the order of $G\approx 100 \, \rm{Pa}$, yielding of the medium has negligible effects on the dynamics of the bubble. These materials have relaxation times that are much larger than typical inverse ultrasonic frequencies, thus, the yielded region behaves as an elastic solid. Conversely, we found that yielding has a significant impact on a bubble oscillating  in the stiff Kaolin suspension. In this case the yielded region behaves as a viscous fluid, which is responsible for an extra oscillation damping. It sets in driving pressure larger than the critical pressure and it induces a sublinear dependence of the oscillation amplitude with the driving pressure.

Our results show that considering the elastic behavior of the yield-stress fluid is crucial to predict yielding of the material and bubble oscillations due to a finite pressure driving. 
Numerical simulations of bubble dynamics in stiff yield-stress fluids suggest that the onset of an additional damping at a critical pressure amplitude $\Delta p \approx \Delta p_{\text{crit}}$ could be exploited to identify the signature of yielding in experiments, which would be cumbersome to assess otherwise. Finally, the numerical and theoretical framework presented in this paper can support experimental investigation of yielding under extensional deformation, which is relatively unexplored compared to yielding under shear deformation.

\section{Acknowledgments} 
This work is supported by European Research Council Starting Grant No. 639221.

\appendix
\setcounter{figure}{0}
\section{Rheological predictions of the EVP model}
\begin{figure}[h!]
\centering
\includegraphics[width=1.0\textwidth]{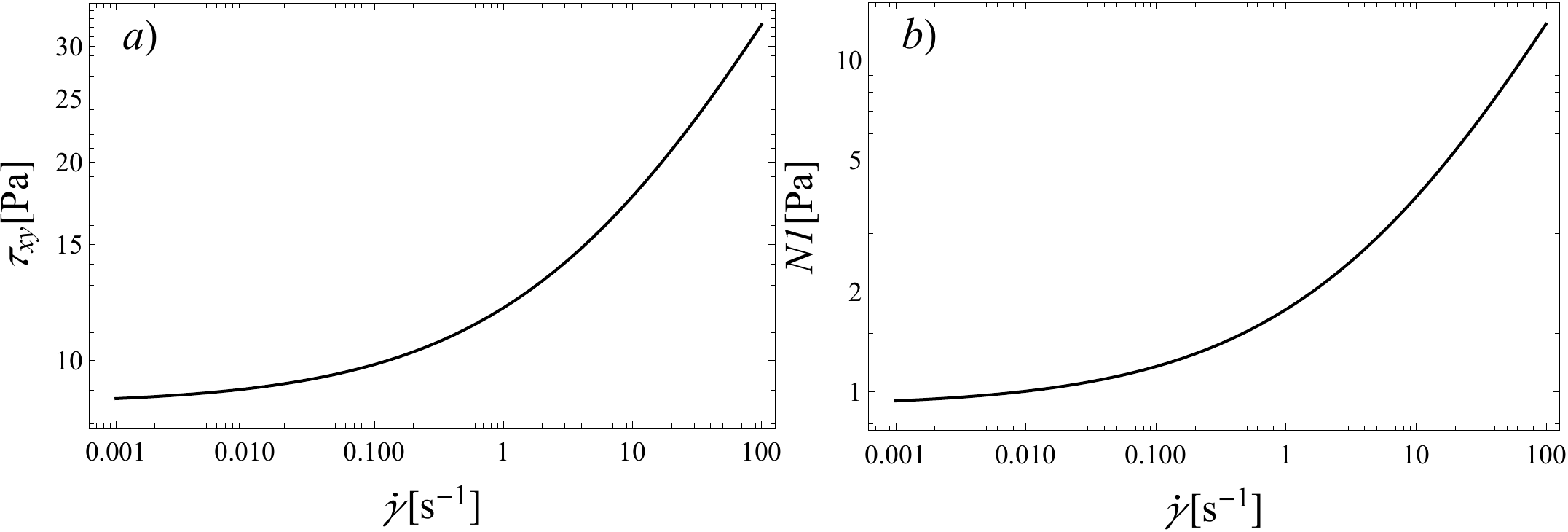}
\caption{a) shear stress, and b) normal stress difference predicted at steady shear by the EVP constitutive model for the Carbopol gel used by Lacaze et al. \cite{lacaze2015steady}. }
\label{fig11}
\end{figure}

\begin{figure}[h!]
\centering
\includegraphics[width=1.0\textwidth]{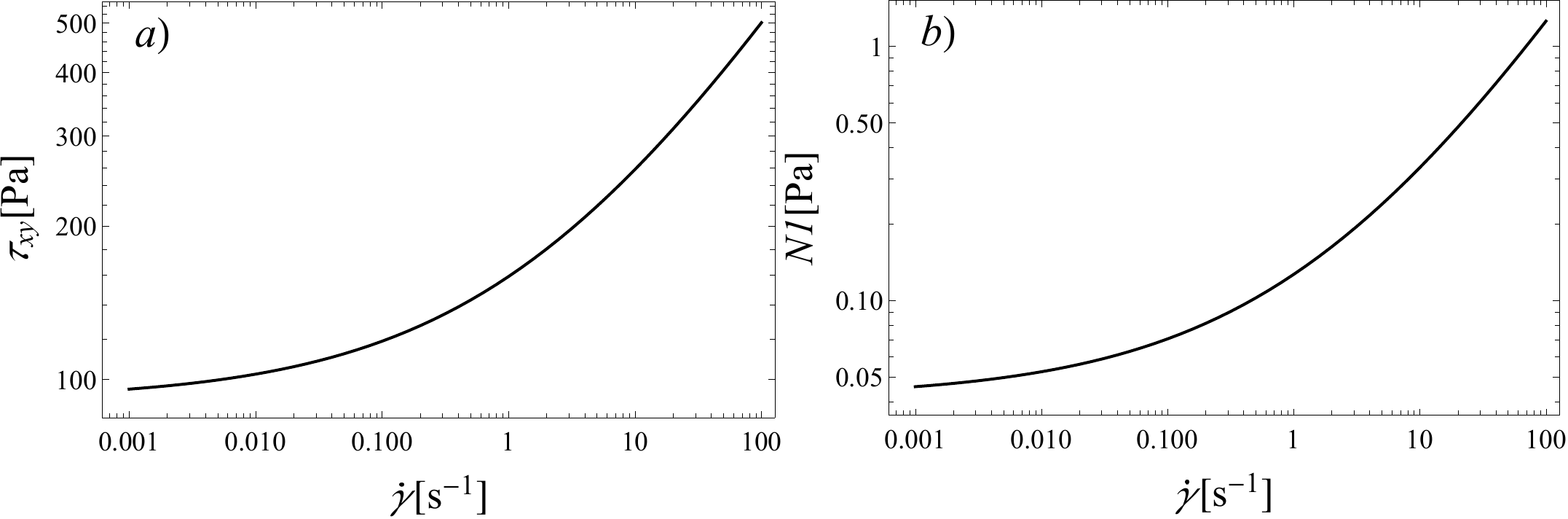}
\caption{a) shear stress, and b) normal stress difference at steady state shear predicted by the EVP constitutive model for the Kaolin suspension employed by Luu and Forterre \cite{luu2009drop}.}
\label{fig12}
\end{figure}

\begin{figure}[h!]
\centering
\includegraphics[width=1.0\textwidth]{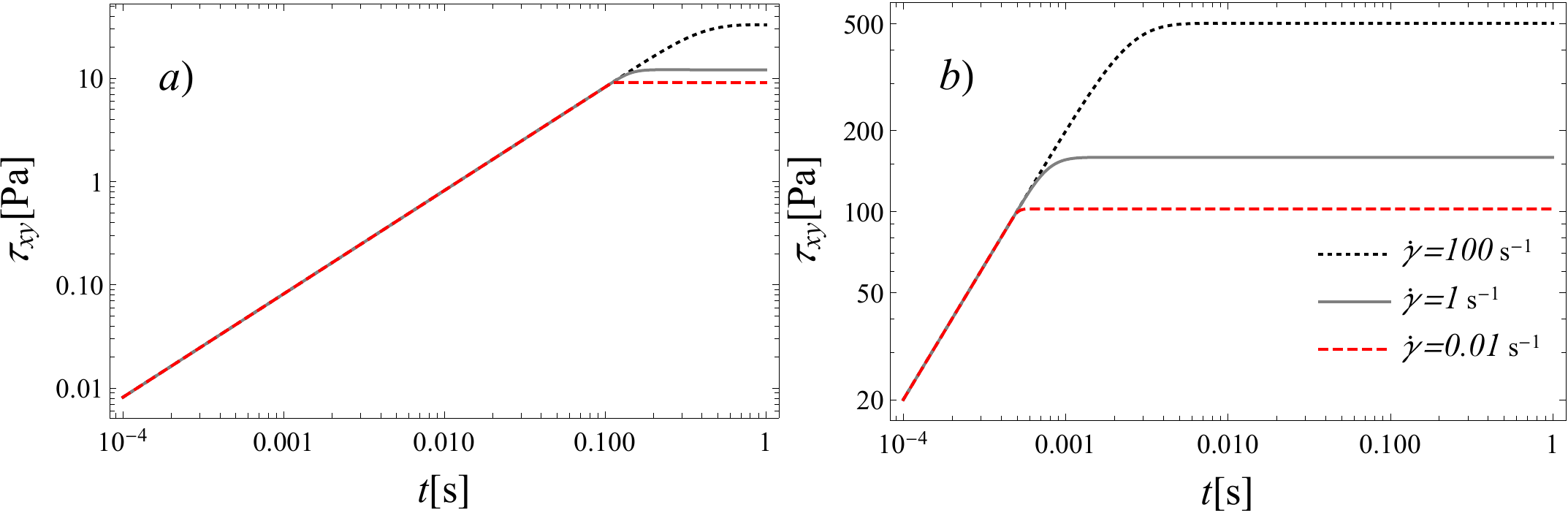}
\caption{Transient shear stress predicted by the EVP constitutive model for a) Carbopol gel and b) Kaolin suspension.}
\label{fig13}
\end{figure}

\begin{figure}[h!]
\centering
\includegraphics[width=1.0\textwidth]{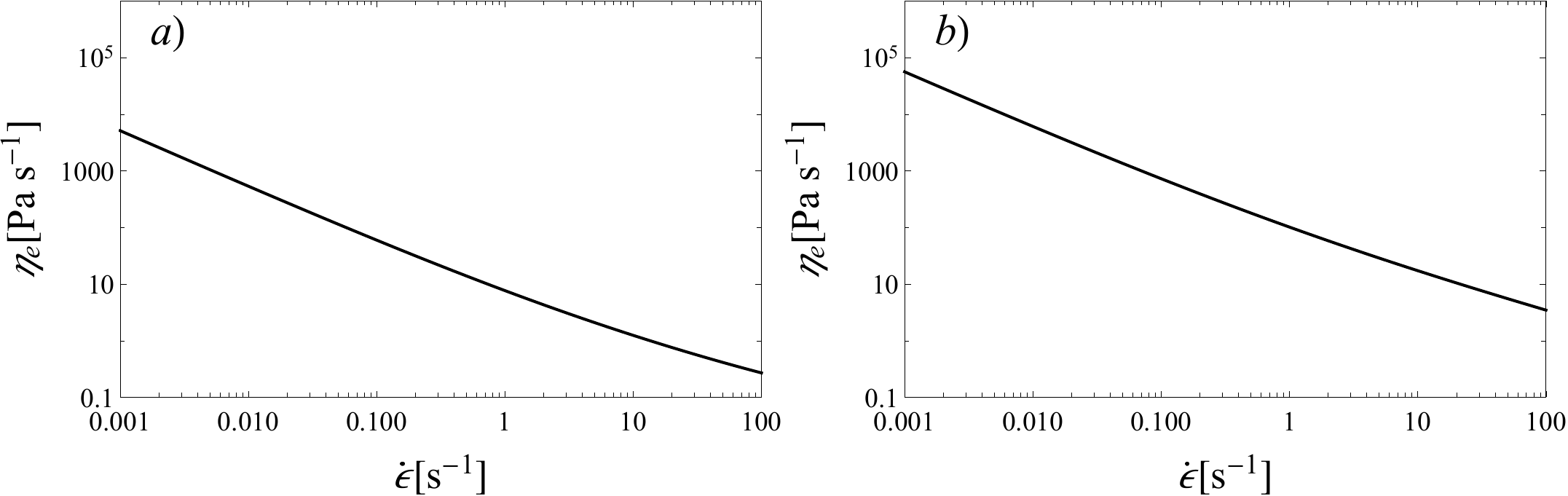}
\caption{Steady state extensional viscosity predicted by the EVP constitutive model for a) Carbopol gel and b) Kaolin suspension.}
\label{fig14}
\end{figure}

In this section, we report the shear and extensional rheology of the Carbopol gel and of the Kaolin suspension predicted by the EVP model. The constitutive parameters are given in Table \ref{table:table1}. We perform shear rheology simulations by fixing the shear rate $\dot{\gamma}$ and computing the shear stress $\tau_{xy}$ and the first normal stress difference $N_1=0.5 (\tau_{xx}-\tau_{yy})$. The second normal stress difference is zero for the EVP constitutive model considered in the present work \cite{fraggedakis2016yielding}. Figures \ref{fig11} and \ref{fig12} show the steady state shear stress and first normal stress difference as a function of the shear rate for the case of the Carbopol gel and the case of the Kaolin suspension, respectively. Both yield-stress fluids considered are shear thinning. The Carbopol gel shows much larger normal stresses than the Kaolin suspension. This behavior is a consequence of its longer relaxation time, which results in larger elastic stresses. The difference in relaxation times between the two yield-stress fluids is better visualized in the transient shear stress response reported in Figure \ref{fig13}. In contrast to the case of the Carbopol gel, the shear stress in the Kaolin suspension reaches its steady state value over $\approx 10^{-3}\, \rm{s}$. 
The extensional rheology simulations are performed by applying a uniaxial extension rate $\dot{\epsilon}$ and computing the steady state extensional viscosity $\eta_e= (\tau_{xx}-\tau_{yy})/3 \dot{\epsilon}$. Figure \ref{fig14}(a) shows that the Carbopol gel is extensional thinning, in agreement with the measurements performed by Louvet et al. \cite{louvet2014nonuniversality}. Figure \ref{fig14}(b) shows that the EVP model predicts extensional thinning also for the case of the Kaolin suspension. Measurements of extensional viscosity of Kaolin suspensions show extensional thinning behavior at low extension rates and extensional thickening at large extension rates \cite{o2002shear}. Since the EVP model predicts extensional thinning (see Fig \ref{fig14}(b)), it is apparent that it is unable to correctly predict the rheological behavior of Kaolin suspensions at large extension rates. Nevertheless, the choice of the EVP constitutive equation to model the Kaolin suspension is justified because we explore extension rates for which experiments report extensional thinning.

\bibliographystyle{elsarticle-num-names} 
\bibliography{Bubble_dynamics_yieldstress}

\end{document}